# Statistics of geometric clusters in Potts model: statistical mechanics approach.

P. N. Timonin

Southern Federal University, 344090, Rostov-on-Don, Stachki ave. 194, Russia.  e-mail: pntim@live.ru

*The percolation of Potts spins with equal values in Potts model on graphs (networks) is considered. The general method for finding the Potts clusters' size distributions is developed. It allows full description of percolation transition when giant cluster of equal-valued Potts spins appears. The method is applied to the short-ranged q-state ferromagnetic Potts model on the Bethe lattices with the arbitrary coordination number z. The analytical results for the field-temperature percolation phase diagram of geometric spin clusters and their size distribution are obtained. The last appears to be proportional to that of the classical non-correlated bond percolation with the bond probability, which depends on temperature and Potts model parameters.*

## 1. Introduction

The thermodynamics of *q*-state Potts model is thoroughly studied on a vast set of graph and lattices [1-3]. In many cases the analytical and numerical results are obtained for its phase diagram, order parameters and critical indexes both for ferromagnetic and antiferromagnetic case. This interest stems from the existence of a number of the model's physical realizations that range from percolation, resistors networks and epidemic spreading [3] to community detection [4]. For these applications more detailed picture and the phase transition mechanisms can be elucidated in the studies of statistics of the like-valued (geometric) spin clusters of Potts model. Geometric clusters distribution is, in principle, observable quantity and its knowledge can be important for the physical realizations of this model. Also, percolation of geometric Potts clusters is very important from the theoretical point of view providing the example of correlated percolation [5].

The influence of geometric clusters percolation in the ferromagnetic spin models on their thermodynamics attracted much attention very early [6-9]. At first, it was suggested that the ferromagnetic transition results from the appearance of infinite percolation cluster of the like-valued spins. However, very soon it was found that such percolation transition may not coincide with the ferromagnetic one [6, 7] and this stimulated further studies of such correlated percolation.

This is notoriously hard problem, which is usually tackled numerically via Monte-Carlo technique [10-14] or via enumeration of clusters in random bonds [15] and sites [16] patterns. Yet such numerics is very time-consuming for sufficiently large samples. The exact analytical results can be obtained with the replica method only for some simple graphs (e.g., various chains and ladders) and without it for the graphs with the tree-like structure.

One can overcome the computational problems via casting percolation into statistical mechanics framework as has been done for the classical bond percolation [17, 18]. This may help to diminish the numerical efforts due to a number of exact methods developed for partition function calculation such as transfer matrix technique, renormalization group and Monte Carlo simulations. Here we suggest the statistical mechanics approach for the determination of size distribution of geometric clusters for Potts model on arbitrary graph.

In section 2, the general method is described for the Potts model on a graph, in Sections 3, 4 the results of its implementation on the Bethe lattice ferromagnetic Potts model are presented and Section 5 is devoted to the discussion and conclusions.

## 2. General formalism

Consider a graph with *N* sites numbered by integers $i = 1,...,N$, and set of bonds *E* between some sites. Potts distribution on graph (non-normalized) is

$$\rho(\boldsymbol{\sigma},h) = \exp\left\{ K \sum_{i<j \in E} \delta(\sigma_i, \sigma_j) + h \sum_i \left[1 - \delta(\sigma_i, 0)\right] \right\}, \quad \sigma_i = \{0,1,...,q-1\} \quad K = J/T, h = H/T \quad (1)$$

$$g(\sigma_i, \sigma_j, \tau_i, \tau_j, \alpha) = 1 + \delta(\sigma_i, \sigma_\alpha) \delta(\sigma_j, \sigma_\alpha) \left[ \delta(\tau_i, \tau_j) - 1 \right], \quad \tau_i = \{0,1,...,\tilde{q}-1\}, \quad (2)$$

$$\sigma_\alpha = \alpha, \quad \alpha = 0,1, \quad g(\sigma_i,\sigma_j,\tau_i,\tau_j,\alpha) = \begin{cases} \delta(\tau_i,\tau_j) & \text{if } \sigma_i = \alpha \text{ and } \sigma_j = \alpha \\ 1 & \text{if } \sigma_i \neq \alpha \text{ or } \sigma_j \neq \alpha \end{cases}$$

Thus, in a given spin configuration all bonds in clusters with all sites having Potts variable $\sigma$ equal to $\alpha$, $\sigma = \alpha$ ($\alpha-clusters$, in short) are endowed with factor $\delta(\tau_i,\tau_j)$. Hence, variables $\tau_i$ have the same values at all sites of every $\alpha-cluster$. Consider partition function

$$Z_{\tilde{q}}(\zeta,h,\alpha) = Tr_{\sigma,\tau} U(\boldsymbol{\sigma},\boldsymbol{\tau},\zeta,h,\alpha), \quad U(\boldsymbol{\sigma},\boldsymbol{\tau},\zeta,h,\alpha) = \rho(\boldsymbol{\sigma},h)\prod_{i,j\in E} g(\sigma_i,\sigma_j,\tau_i,\tau_j,\alpha)\prod_{i=1}^{N}\zeta^{(1-\delta_{\tau_i,\alpha})}.$$

In Fortuin-Kasteleyn cluster representation [17] it has the form

$$Z_{\tilde{q}}(\zeta,h,\alpha) = Tr_{\boldsymbol{\sigma}}\left\{\rho(\boldsymbol{\sigma},h)\left[\sum_{\tau=0}^{\tilde{q}-1}\zeta^{(1-\delta_{\tau,\alpha})}\right]^{\sum_{\beta\neq\alpha}N_\beta(\boldsymbol{\sigma})}\prod_{s=1}^{N}\left[\sum_{\tau=0}^{\tilde{q}-1}\zeta^{s(1-\delta_{\tau,\alpha})}\right]^{N_s^{(\alpha)}(\boldsymbol{\sigma})}\right\} = $$
$$= Tr_{\boldsymbol{\sigma}}\left\{\rho(\boldsymbol{\sigma},h)\left[1+(\tilde{q}-1)\zeta\right]^{\sum_{\beta\neq\alpha}N_\beta(\boldsymbol{\sigma})}\prod_{s=1}^{N}\left[1+(\tilde{q}-1)\zeta^s\right]^{N_s^{(\alpha)}(\boldsymbol{\sigma})}\right\}$$ (3)

where $N_s^{(\alpha)}(\boldsymbol{\sigma})$ is the number of clusters with $s$ sites, each having $\sigma = \alpha$ in configuration $\boldsymbol{\sigma}$. Apparently,

$$\left\langle\frac{\sum_{\beta\neq\alpha}N_\beta(\boldsymbol{\sigma})}{N}\right\rangle_P = 1 - \left\langle\frac{N_\alpha(\boldsymbol{\sigma})}{N}\right\rangle_P$$

At $\tilde{q}-1$ small we have approximately

$$Z_{\tilde{q}}(\zeta,h,\alpha) \approx Z_1(\zeta,h,\alpha)\left[1+(\tilde{q}-1)\sum_{s=1}^{N}\zeta^s\left\langle\frac{N_s^{(\alpha)}(\boldsymbol{\sigma})}{N}\right\rangle_P + (\tilde{q}-1)\zeta\left(1-\left\langle\frac{N_\alpha(\boldsymbol{\sigma})}{N}\right\rangle_P\right)\right],$$ (4)

Further we note that at $\tilde{q}=1$ partition function (4) becomes the ordinary Potts one on a graph considered

$$Z_1(\zeta,h,\alpha) = Tr_{\boldsymbol{\sigma}}\rho(\boldsymbol{\sigma},h) \equiv Z_{Potts}(h),$$

and the skew brackets with sub-index $P$ denote ordinary Potts averages

$$\langle A(\boldsymbol{\sigma})\rangle_P \equiv Z_P^{-1}(h)Tr_{\boldsymbol{\sigma}}\left[\rho(\boldsymbol{\sigma},h)A(\boldsymbol{\sigma})\right].$$

Let us define the thermodynamic potential

$$F_{\tilde{q}}(\zeta,h,\alpha) = \lim_{N\to\infty} N^{-1} \ln Z_{\tilde{q}}(\zeta,h,\alpha),$$ (5)

Note that at $\tilde{q}=1$ $F_{\tilde{q}}(\zeta,h,\alpha)$ becomes proportional to the ordinary Potts potential

$$F_1(\zeta,h,\alpha) = -\beta F_{Potts}(h)$$

Then their difference

$$\Phi_{\tilde{q}}(\zeta,h,\alpha) = F_{\tilde{q}}(\zeta,h,\alpha) - F_1(h)$$ (6)

define the generating function for clusters' size distribution

$$G_\alpha(\zeta,h) = \sum_{s=1}^{\infty} \zeta^s v_s^{(\alpha)}, \qquad v_s^{(\alpha)} = \lim_{N\to\infty} \left\langle \frac{N_s^{(\alpha)}(\boldsymbol{\sigma})}{N} \right\rangle_P \qquad (7)$$

Here $v_s^{(\alpha)}$ is the average number (per site) of $\alpha-clusters$ with $s$ sites (composed of sites with $\sigma_i = \alpha$). Indeed, we have from (4)

$$G_\alpha(\zeta,h) = \partial_{\tilde{q}} \Phi_{\tilde{q}}(\zeta,h,\alpha)\big|_{\tilde{q}=1} - \zeta(1-c_\alpha), \qquad c_\alpha = \lim_{N\to\infty} \left\langle \frac{N_\alpha(\boldsymbol{\sigma})}{N} \right\rangle_P \qquad (8)$$

Now we only need to find $c_\alpha$ - the average fraction of sites with $\sigma = \alpha$. We can express them via average Potts spin

$$\lim_{N\to\infty} N^{-1} \left\langle \sum_{i=1}^{N} \sigma_i \right\rangle_P = \langle \sigma_i \rangle_P = \sum_{\beta=0}^{q-1} \beta c_\beta$$

Due to the permutation symmetry of Potts spins with $\sigma_i \neq 0$ $c_\alpha = c_1$ for all $\alpha > 0$ so

$$\langle \sigma \rangle_P = c_1 \sum_{\beta=0}^{q-1} \beta = \frac{c_1}{2} q(q-1)$$

By definition we also have $\sum_{\beta=0}^{q-1} c_\beta = c_0 + (q-1)c_1 = 1$. Thus we have

$$c_0 = 1 - \frac{2}{q} \langle \sigma_i \rangle_P, \qquad c_1 = \frac{2}{q(q-1)} \langle \sigma_i \rangle_P$$

Note also that $c_\alpha$ can be expressed via Potts "magnetization" conjugate to field $h$

$$m_P = \frac{\partial F_1}{\partial h} = \langle 1 - \delta(\sigma_i, 0) \rangle_P = 1 - c_0 = (q-1)c_1 \qquad (9)$$

Once we have $G_\alpha(\zeta,h)$ from (8) we can find the average number of finite $\alpha-clusters$

$$n_{cl,\alpha} = \sum_{s=1}^{\infty} v_s^{(\alpha)} = G_\alpha(1,h),$$

and the average number of sites in them

$$n_{\alpha-sites} = \sum_{s=1}^{\infty} s v_s^{(\alpha)} = \partial_\zeta G_\alpha(\zeta,h)\big|_{\zeta=1}$$

Let $P_\alpha$ to be the fraction of $\alpha-sites$ belonging to the infinite percolation $\alpha-cluster$ then

$$P_\alpha = c_\alpha - n_{\alpha-sites}$$

Thus we get full description of the Potts clusters percolation on a graph if we manage to calculate the potential $\Phi_{\tilde{q}}(\zeta,h,\alpha)$ (6).

## 3. Application to a Bethe lattice

The Bethe lattice is a popular example of hierarchical graph for which a wealth of analytical results on phase transitions in spin and percolation models were obtained [3, 6, 8, 19, 20]. Here we apply the above method to find analytically the geometric clusters' size distribution for Potts model on a general Bethe lattice with a coordination number

$z$. The potential (5) for it can be found with the partial partition function for the Caley tree of $l$-th order $V_{\alpha,l}(\sigma,\tau)$ summed over all spins except the root ones $(\sigma,\tau)$ [19]

$$F_{\tilde{q}}(\zeta,h,\alpha) = \lim_{N\to\infty} N^{-1} \ln Z_{\tilde{q}}(\zeta,h,\alpha) = \frac{2-z}{2} \lim_{l\to\infty} \ln Tr_{\sigma,\tau} e^{h[1-\delta_{\sigma,0}]} \zeta^{1-\delta_{\tau,\alpha}} V_{\alpha,l}^z(\sigma,\tau). \qquad (10)$$

$V_{\alpha,l}(\sigma,\tau)$ obeys the recursion relations

$$V_{\alpha,l+1}(\sigma,\tau) = \sum_{\sigma'} e^{K\delta(\sigma,\sigma')+h[1-\delta(\sigma',0)]} \sum_{\tau'} g(\sigma,\tau,\sigma',\tau',\alpha) \zeta^{1-\delta(\tau',\alpha)} W_{\alpha,l}(\sigma',\tau'), \quad W_{\alpha,l}(\sigma,\tau) = \left[V_{\alpha,l}(\sigma,\tau)\right]^{z-1} \qquad (11)$$

so with its stationary values at $l\to\infty$ $V_{\alpha,\infty}(\sigma,\tau) \equiv V_\alpha(\sigma,\tau)$ we get the potential for $\alpha=0,1$ using its symmetry under permutations of $\sigma\neq 0$ and $\tau\neq\alpha$

$$F_{\tilde{q}}(\zeta,h,\alpha) = \frac{2-z}{2} \ln \left\{ \begin{array}{l} V_\alpha(0,\alpha)W_\alpha(0,0) + (q-1)e^h V_\alpha(1,\alpha)W_\alpha(1,\alpha) + \\ +(\tilde{q}-1)\zeta\left[V_\alpha(0,1-\alpha)W_\alpha(0,1-\alpha) + (q-1)e^h V_\alpha(1,1-\alpha)W_\alpha(1,1-\alpha)\right] \end{array} \right\}, \qquad (12)$$

$$W_\alpha(\sigma,\tau) \equiv \left[V_\alpha(\sigma,\tau)\right]^{z-1}$$

Summing over spins in (8) we get the stationary values' equations for $\alpha=0,1$

$$V_\alpha(\sigma,\tau) = W_\alpha(0,\alpha) + (\tilde{q}-1)\zeta W_\alpha(0,1-\alpha) + (q-1)e^h\left[W_\alpha(1,\alpha) + (\tilde{q}-1)\zeta W_\alpha(1,1-\alpha)\right] +$$
$$+ (e^K - 1)e^{h[1-\delta(\sigma,0)]}\left[W_\alpha(\sigma,\alpha) + (\tilde{q}-1)\zeta W_\alpha(\sigma,1-\alpha)\right] + \qquad (13)$$
$$+ e^K \delta(\sigma,\alpha) e^{h[1-\delta(\alpha,0)]} \left\{ \zeta^{1-\delta(\tau,\alpha)} W_\alpha(\alpha,\tau) - \left[W_\alpha(\alpha,\alpha) + (\tilde{q}-1)\zeta W_\alpha(\alpha,1-\alpha)\right] \right\}$$

Apparently $V_\alpha(\sigma,\alpha) - V_\alpha(\sigma,1-\alpha) = e^K \delta(\sigma,\alpha) e^{h[1-\delta(\sigma,0)]} \left[W_\alpha(\alpha,\alpha) - \zeta W_\alpha(\alpha,1-\alpha)\right]$

### 3.1. Case $\alpha = 0$

Introducing variables

$$v = \frac{W_0(0,0)}{W_0(1,0)}, \quad u = \frac{W_0(0,1)}{W_0(1,0)},$$

we can express potential $F_{\tilde{q}}(\zeta,h,0)$ via just these two variables, see Appendix A.

$u$ and $v$ obey the equations (A1) that follow from (13). The solutions to them at small $\varepsilon \equiv \tilde{q}-1$ can be represented as

$$u \approx e^h(y + \varepsilon \zeta y_1), \quad v \approx e^h(x + \varepsilon \zeta x_1). \qquad (14)$$

From (A1) the equations for $x$ and $y$ follow

$$x = e^{-h} \left[\frac{e^K x + q - 1}{x + e^K + q - 2}\right]^{z-1}, \qquad (15)$$

$$y = e^{-h} \left[\frac{q - 1 + e^K \zeta y}{x + e^K + q - 2}\right]^{z-1}, \qquad (16)$$

while $x_1$ is expressed via $x$ and $y$

$$x_1 = (z-1)x \frac{x(e^K+q-1)(1-e^K)-y(e^K x+q-1)}{(e^K x+q-1)(x+e^K+q-2)+(z-1)(e^K+q-1)(1-e^K)x}, \quad (17)$$

Substituting (15-17) into (A2) we get in first order in $\varepsilon \equiv \tilde{q}-1$

$$F_{\tilde{q}}(h,\zeta,\alpha=0) \approx F_1(h,\alpha=0)+\varepsilon G_0(w)+\varepsilon\zeta(1-c_0)$$

where

$$G_0(w) = c_0 \frac{(q-1)w+\frac{2-z}{2}e^K w^2}{x(e^K x+q-1)}, \quad w=\zeta y \quad (18)$$

$$F_1(\alpha=0,x) = h+(z-1)\ln\left[x+e^K+q-2\right]+\frac{2-z}{2}\ln\left[e^K x^2+2(q-1)x+(q-1)(e^K+q-2)\right] \quad (19)$$

$$c_0(x) = 1-m_P(x) = \frac{x(e^K x+q-1)}{e^K x^2+2(q-1)x+(q-1)(e^K+q-2)}, \quad (20)$$

$$m_P(x) = \frac{\partial F_1(x,\alpha=0)}{\partial h} = \frac{\partial F_1(x,\alpha=0)}{\partial x}\frac{\partial x}{\partial h} = \frac{(q-1)(x+e^K+q-2)}{e^K x^2+2(q-1)x+(q-1)(e^K+q-2)}, \quad (21)$$

$w=w(\zeta)$ obeys the equation

$$w = e^{-h}\zeta\left[\frac{q-1+e^K w}{x+e^K+q-2}\right]^{z-1}, \quad (22)$$

so

$$\frac{\partial w}{\partial \zeta} = \frac{w}{\zeta}\frac{q-1+e^K w}{q-1+(2-z)e^K w} \quad (23)$$

and we have

$$n_{0-sites} = \partial_\zeta G_0[w(\zeta)]\big|_{\zeta=1} = \partial_w G_0[w]\frac{\partial w}{\partial \zeta}\big|_{\zeta=1} = c_0 \frac{w_1(e^K w_1+q-1)}{x(e^K x+q-1)}, \quad n_{cl,0} = G_0(w_1). \quad (24)$$

$w_1$ obeys the equation

$$w_1 = e^{-h}\left[\frac{q-1+e^K w_1}{x+e^K+q-2}\right]^{z-1} \quad (25)$$

For the capacity of giant 0-cluster we have

$$P_0 = c_0 - n_{0-sites} = c_0 - c_0\frac{w_1(e^K w_1+q-1)}{x(e^K x+q-1)} = c_0\frac{(x-w_1)[q-1+e^K(x+w_1)]}{x(e^K x+q-1)} \quad (26)$$

Thus we have obtained the thermodynamic parameters (19-21) and the percolation ones (19), (24), (26) as functions of x (15), w (22) and $w_1$ (25). Finding stable solutions of (15), (22), (25) and substituting them into the expressions for the thermodynamic and the percolation parameters we can obtain their values at all $T/J=K^{-1}$ and $H/J=hK^{-1}$. Note also that above equations present the parametric representations of the thermodynamic and the percolation parameters as functions of $T/J$ and $H/J$ which greatly simplifies their graphing.

Most simply 0-clusters' size distribution $v_s^{(0)}$ can be obtained from (18), (22) using the change of integration variable and the integration by parts

$$v_s^{(0)} = \oint_{|\zeta|=c} \frac{d\zeta}{2\pi i} \frac{G_0[w(\zeta)]}{\zeta^{s+1}} = \oint_{|w|=\rho} \frac{dw}{2\pi i s} \zeta^{-s}(w) \frac{dG_0(w)}{dw} \qquad (27)$$

and the relation

$$\oint_{|w|=\rho} \frac{dw}{w^m}(w+R)^n = \frac{\partial_w^{m-1}}{(m-1)!}(w+R)^n \bigg|_{w=0} = \frac{\partial_R^{m-1} R^n}{(m-1)!} = \binom{n}{m-1} R^{n-m+1}.$$

So we get

$$v_s^{(0)} = c_0 \frac{z}{st} p^{s-1}(1-p)^t \binom{s(z-1)}{s-1}, \qquad (28)$$

where

$$p = \frac{e^K x}{e^K x + q - 1}, \quad t = s(z-2) + 2. \qquad (29)$$

Here $t$ is the number of empty bonds in the perimeter of $s$-site 0-cluster and $b = s-1$ is the number of bonds inside it. Thus $v_s^{(0)}$ in (28) coincides up to pre-factor $c_0$ with the sizes' distribution of clusters that appear in the process of the independent random placement the bonds with probability $p$ [19], i.e., in the classical bond percolation.

### 3.2. Case $\alpha \neq 0$

Similarly to previous case, we introduce the variables

$$\tilde{u} = \frac{W_1(1,0)}{W_1(0,0)}, \quad \tilde{v} = \frac{W_1(1,1)}{W_1(0,0)},$$

which obey the equations

$$\tilde{u} = \left\{ \frac{[1+(\tilde{q}-1)\zeta] + (q-2)e^h[\tilde{v}+(\tilde{q}-1)\zeta\tilde{u}] + e^K e^h \zeta\tilde{u}}{e^K[1+(\tilde{q}-1)\zeta] + (q-1)e^h[\tilde{v}+(\tilde{q}-1)\zeta\tilde{u}]} \right\}^{z-1},$$

$$\tilde{v} = \left\{ \frac{[1+(\tilde{q}-1)\zeta] + (q-2)e^h[\tilde{v}+(\tilde{q}-1)\zeta\tilde{u}] + e^K e^h \tilde{v}}{e^K[1+(\tilde{q}-1)\zeta] + (q-1)e^h[\tilde{v}+(\tilde{q}-1)\zeta\tilde{u}]} \right\}^{z-1},$$

Again, the thermodynamic potential $F_{\tilde{q}}(h,\zeta,\alpha=1)$ can be expressed as function of these two variables.

At small $\varepsilon \equiv \tilde{q}-1$

$$\tilde{u} \approx e^{-h}(\tilde{y} + \varepsilon \zeta \tilde{y}_1), \quad \tilde{v} \approx e^{-h}(\tilde{x} + \varepsilon \zeta \tilde{x}_1),$$

where $\tilde{x}$ and $\tilde{y}$ obey the equations

$$\tilde{x} = e^h \left\{ \frac{1+(q-2+e^K)\tilde{x}}{e^K+(q-1)\tilde{x}} \right\}^{z-1}, \quad \tilde{y} = e^h \left\{ \frac{1+(q-2)x+e^K\zeta\tilde{y}}{e^K+(q-1)\tilde{x}} \right\}^{z-1}, \quad (30)$$

Using similar procedure, we obtain

$$F_{\tilde{q}}(h,\zeta,\alpha=1) \approx F_1(\tilde{x},\alpha=1) + \varepsilon G_1(\zeta) + \varepsilon\zeta(1-c_1)$$

$$F_1(\alpha=1,\tilde{x}) = (z-1)\ln\left[e^K+(q-1)\tilde{x}\right] + \frac{2-z}{2}\ln\left[e^K+2(q-1)\tilde{x}+(e^K+q-2)(q-1)\tilde{x}^2\right], \quad (31)$$

$$G_1(\zeta) = \zeta(c_1-1+c_0) + \tilde{G}_1\left[\tilde{w}(\zeta)\right] = \zeta(2-q)c_1 + \tilde{G}_1\left[\tilde{w}(\zeta)\right],$$

$$\tilde{G}_1(\tilde{w}) = m_P(\tilde{x})\tilde{w} \frac{1+(q-2)\tilde{x}+\frac{2-z}{2}e^K\tilde{w}}{\tilde{x}\left[1+(q-2+e^K)\tilde{x}\right]}$$

$$m_P(\tilde{x}) = \frac{\partial F_1}{\partial h} = \frac{\partial F_1(\tilde{x},\alpha=1)}{\partial \tilde{x}} \frac{\partial \tilde{x}}{\partial h} = \frac{(q-1)\tilde{x}\left[1+(q-2+e^K)\tilde{x}\right]}{(q-1)(e^K+q-2)\tilde{x}^2 + 2(q-1)\tilde{x}+e^K} \quad (32)$$

$$c_1(\tilde{x}) = \frac{m_P(\tilde{x})}{q-1} = \frac{\tilde{x}\left[1+(q-2+e^K)\tilde{x}\right]}{(q-1)(e^K+q-2)\tilde{x}^2 + 2(q-1)\tilde{x}+e^K}$$

$$\tilde{w} = \zeta\tilde{y}, \quad \tilde{w} = \zeta e^h \left(\frac{1+(q-2)\tilde{x}+e^K\tilde{w}}{e^K+(q-1)\tilde{x}}\right)^{z-1}, \quad \frac{\partial \tilde{w}}{\partial \zeta} = \frac{\tilde{w}}{\zeta} \frac{1+(q-2)\tilde{x}+e^K\tilde{w}}{1+(q-2)\tilde{x}+(2-z)e^K\tilde{w}},$$

$$\tilde{w}_1 = e^h \left(\frac{1+(q-2)\tilde{x}+e^K\tilde{w}_1}{e^K+(q-1)\tilde{x}}\right)^{z-1}, \quad (33)$$

$$n_{cl,1} = G_1(\zeta=1) = (2-q)c_1 + \tilde{G}_1(\tilde{w}_1), \quad n_{1-sites} = \partial_\zeta G_1(\zeta)\big|_{\zeta=1} = (2-q)c_1 + \partial_{\tilde{w}}\tilde{G}_1(\tilde{w})\big|_{\tilde{w}=\tilde{w}_1} \frac{\partial \tilde{w}}{\partial \zeta}\bigg|_{\zeta=1}, \quad (34)$$

$$n_{1-sites} = (2-q)c_1 + m_P(\tilde{x}) \frac{\tilde{w}_1}{\tilde{x}} \frac{1+(q-2)\tilde{x}+e^K\tilde{w}_1}{1+(q-2+e^K)\tilde{x}} \quad (35)$$

$$P_1 = c_1 - n_{1-sites} = (q-1)c_1 - m_P(\tilde{x}) \frac{\tilde{w}_1}{\tilde{x}} \frac{1+(q-2)\tilde{x}+e^K\tilde{w}_1}{1+(q-2+e^K)\tilde{x}} = m_P \frac{\tilde{x}-\tilde{w}_1}{\tilde{x}} \frac{1+(q-2)\tilde{x}+e^K(\tilde{w}_1+\tilde{x})}{1+(q-2+e^K)\tilde{x}} \quad (36)$$

As before, we need to solve the equations for $\tilde{x}$ (30) and $\tilde{w}_1$ (33) for stable solutions and substitute them to get the final physical results from (34-36).

For 1-clusters` size distribution we get similarly

$$v_{s>1}^{(1)} = m_P \frac{z}{st}\binom{s(z-1)}{s-1}\tilde{p}^{s-1}(1-\tilde{p})^t, \quad v_1^{(1)} = c_1\left[2-q+(q-1)(1-\tilde{p})^z\right] \quad (37)$$

$$\tilde{p} = \frac{e^K\tilde{x}}{1+(q-2+e^K)\tilde{x}}, \quad t = s(z-2)+2. \quad (38)$$

Same as $v_s^{(0)}$ in (28), $v_s^{(1)}$ coincides up to pre-factor $m_P = 1 - c_0$ with the sizes' distribution of clusters in the classical bond percolation with probability $\tilde{p}$ (38).

## 4. Thermodynamic and percolation phase diagram.

Let us first consider the relation between the expressions for Potts thermodynamic variables that are obtained in Sections 3.1 and 3.2. Here we have two distinct sets of formulas for the same parameters – magnetization $m_P$, $c_\alpha$ and $F_1(\alpha)$. Actually, they are the same as it should be. To see this we note that equation $x^{-1}(h) = \tilde{x}(h)$ follows from (15) and (30). Then inspecting (21) and (32), we find that they transform one to another under the change $x(h) \to x^{-1}(h)$

$$m_P[\alpha = 0, x(h)] = m_P[\alpha = 1, x^{-1}(h)]$$

so they have equal values as $x^{-1}(h) = \tilde{x}(h)$. The same goes for $c_\alpha$. For the thermodynamic potential we have from (19) and (31)

$$F_1(\alpha = 0, x) = h + (z-1)\ln(x + e^K + q - 2) + \frac{2-z}{2}\ln\left[e^K x^2 + 2(q-1)x + (q-1)(e^K + q - 2)\right] =$$

$$= (z-1)\ln(e^K x + q - 1) - \ln x + \frac{2-z}{2}\ln\left[e^K x^2 + 2(q-1)x + (q-1)(e^K + q - 2)\right] =$$

$$= (z-1)\ln\left[e^K + (q-1)x^{-1}\right] + \frac{2-z}{2}\ln\left[e^K + 2(q-1)x^{-1} + (q-1)(e^K + q - 2)x^{-2}\right] = F_1(\alpha = 1, x^{-1})$$

Here we used the equation of state (15) according to which

$$h(x) = (z-1)\ln\frac{e^K x + q - 1}{e^K + x + q - 2} - \ln x. \qquad (39)$$

Hence

$$F_1(\alpha = 0, x) = F_1(\alpha = 1, \tilde{x}) = -\beta F_{Potts},$$

so we can use both equivalent representations to describe thermodynamics of the Potts model.

Furthermore, we choose the $\alpha = 0$ representation to consider the influence of thermodynamics on the percolation of the geometric $\alpha - clusters$.

Stable solutions to the equation of state (15) must obey the condition

$$e^{-h}\frac{d}{dx}\left[\frac{e^K x + q - 1}{e^K + x + q - 2}\right]^{z-1} = (z-1)x\frac{(e^K - 1)(e^K + q - 1)}{(e^K x + q - 1)(e^K + x + q - 2)} < 1, \qquad (40)$$

This is quadratic inequality with respect to $x$. It fulfils at all $x$ if

$$R(z, q, T) = (e^{J/T} - 1)(e^{J/T} + q - 1)\left[(z-2)^2(e^{J/T} - 1)(e^{J/T} + q - 1) - 4(z-1)(q-1)\right] < 0. \qquad (41)$$

When $R(z, q, T) > 0$, that is at $T < T_t(z, q)$

$$T_t(z, q) = J / \ln\frac{\sqrt{q^2 z^2 - 4(z-1)(q-2)^2} - (z-2)(q-2)}{2(z-2)} \qquad (42)$$

(40) holds for

$$x_-(z,q,T) < x < x_+(z,q,T), \qquad (43)$$

$$x_s(z,q,T) = \frac{1}{2e^{J/T}}\left[(z-2)(e^{J/T}-1)(e^{J/T}+q-1)-2(q-1)+s\sqrt{R(z,q,T)}\right]. \qquad (44)$$

There are two stable solutions to the equation of state (15) at $T < T_t(z,q)$ that obey the condition (43). On the *H-T* plane the region where they coexist lies at

$$H_-(z,q,T) < H < H_+(z,q,T), \qquad (45)$$

where the expressions for $H_s(z,q,T)$ follow from (39)

$$H_s(z,q,T)/T = h[x_s(z,q,T)] = (z-1)\ln\left[\frac{e^{J/T}x_s(z,q,T)+q-1}{e^{J/T}+x_s(z,q,T)+q-2}\right] - \ln x_s(z,q,T), \qquad (46)$$

It follows from (46) that $H_s(z,q,T=0) = s(z-2)$.

At the field

$$H_{tr}(z,q,T) = T(z-1)\ln\left(\frac{e^K x_{tr}+q-1}{x_{tr}+e^K+q-2}\right) - T\ln x_{tr} = \frac{T}{2}(z-2)\ln(q-1) - \frac{T}{2}z\ln\left[1+(q-2)e^{-J/T}\right] \qquad (47)$$

the first order transition takes place between the equilibrium phase with $m_P < \frac{1}{2}$ (having the lowest potential at $H < H_{tr}(z,q,T)$) into another equilibrium at $H > H_{tr}(z,q,T)$ phase with $m_P > \frac{1}{2}$.

The equation (47) follows from the Maxwell rule [20], see Appendix B.

The dashed line in Figure 1 divides the *H-T* plane into two regions in which $m_P < \frac{1}{2}$ and $m_P > \frac{1}{2}$ in equilibrium states. Note that at all *T* this line is described by (47) but only at $T < T_t(z,q)$ it designates the first order transition.

At $T = T_t(z,q)$

$$H_-(z,q,T) = H_+(z,q,T) = H_{tr}(z,q,T)$$

so all three lines meet here. Their meeting point is called the tricritical point, its coordinates on *H-T* plane are

$$\{T_t(z,q), H_t(z,q) = H_t[z,q,T_t(z,q)]\}. \qquad (48)$$

Beside the equilibrium phases in the region (45) there are two metastable ones - with $m_P > \frac{1}{2}$ at $H < H_{tr}(z,q,T)$ and with $m_P < \frac{1}{2}$ at $H > H_{tr}(z,q,T)$. At the boundaries of the coexistence region (45) susceptibility $\chi = \frac{\partial m_P}{\partial h}$ diverges in these states.

The transition line (47) always crosses the line $H = 0$ and there always exists the first order transition at $H = 0$ at which $m_P$ drops down at

$$T_{tr}(z,q) = J/\ln\frac{q-2}{(q-1)^{\frac{z-2}{z}}-1} \qquad (49)$$

Note also that the lower boundary of the coexistence region (45) always touch the line H = 0 at $T = T_c(z,q)$,

$$H_-\left[z,q,T_c(z,q)\right]=0 \;,\; T_c(z,q) = J / \ln \frac{q+z-2}{z-2} \qquad (50)$$

This coexistence region (45) is shown in Figure 1. Figure 2 shows the field dependence of magnetization $m_P$ and susceptibility $\chi = \dfrac{\partial m_P}{\partial h}$. Figure 3 shows the temperature dependency of $m_P$ at H = 0.

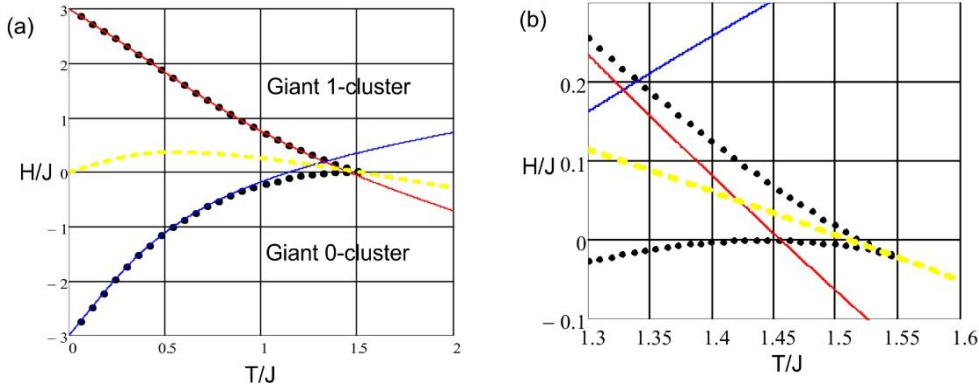

Figure 1. (a) *H-T* phase diagram for Potts model on Bethe lattice with *q*=3, *z*=5, $T_t$ = 1.546J, $H_t$ = -0.02J, $T_{tr}$= 1.51J, $T_c$ =1.443J. Dotted lines denote coexistence region (45), dashed line – first order transition, above upper full (red) line giant 1-cluster exist, under lower (blue) line giant 0-cluster exists. (b) the vicinity of tricritical point expanded.

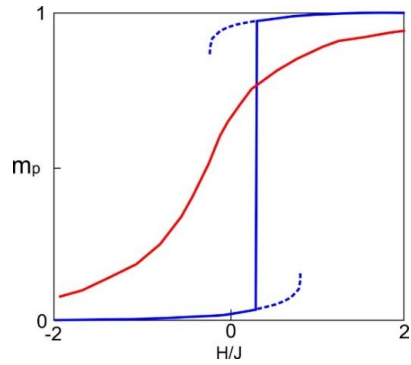

Figure 2. Field dependence of $m_P = \dfrac{\partial F_{Potts}}{\partial h}$ at *T=J* and *T=2J*. Dashed lines correspond to metastable states.

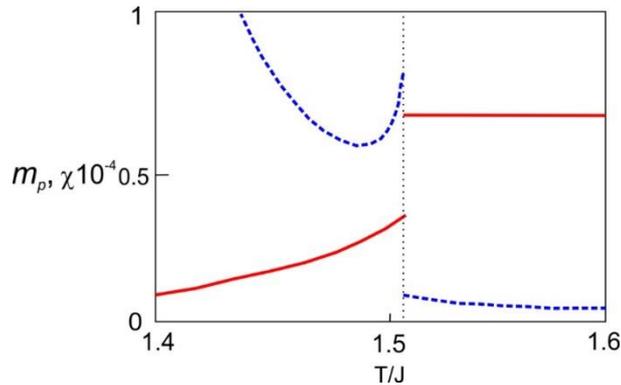

Figure 3. Temperature dependencies of $m_P = \dfrac{\partial F_{Potts}}{\partial h}$ (full line) and $\chi = \dfrac{\partial m_P}{\partial h}$ (dashed line) at H=0.

We should note that above formulae (42-50) coincide with that of [21] for ferromagnetic Potts model on Bethe lattice after due account of the relation of our parameters to that of [21]. In this paper authors add to the Potts Hamiltonian the term $-H'\sum_i \delta(\sigma_i,1)$ so our $h = \dfrac{H}{T} = -H'$ and the magnetization in this paper

$M = -\dfrac{\partial \beta F_{Potts}}{\partial H'} = \langle \delta(\sigma_i,1) \rangle$ is equal to our $c_0$.

The stable solutions to the percolation equation of state for the 0-clusters (25) must obey the condition

$e^{-h} \dfrac{d}{dw_1}\left[ \dfrac{q-1+e^K w_1}{x+e^K+q-2}\right]^{z-1} = w_1 e^K \dfrac{z-1}{q-1+e^K w_1} < 1$ or $w_1 < \dfrac{q-1}{e^K(z-2)} \equiv x_{p0}$. On the other hand the infinite 0-cluster does not exist in phases where $w_1 > x$, see (26). Thus, non-percolating phase exists when $x < w_1 < x_{p0}$ and percolation transition takes place at $x = x_{p0}$ or at

$$H_{p0} = -J(z-2) + T\left\{(z-1)\ln\left[\dfrac{(q-1)(z-1)}{(z-2)\left[1+e^{-J/T}(q-2)\right]+e^{-2J/T}(q-1)}\right] - \ln\dfrac{q-1}{z-2}\right\} \quad (51)$$

In the stable thermodynamic phases (equilibrium and metastable) $\dfrac{\partial x}{\partial h} < 0$ so the non-percolating condition $x < x_{p0}$ means that the infinite 0-cluster is absent at $H_{p0} < H$ while it is present at $H < H_{p0}$.

Meanwhile the stable solutions to the percolation equation of state for the 1-clusters (33) must obey the condition $e^h \dfrac{\partial}{\partial \tilde{w}_1}\left(\dfrac{1+(q-2)\tilde{x}+e^K \tilde{w}_1}{e^K+(q-1)\tilde{x}}\right)^{z-1} < 1$ or $\tilde{w}_1 < \dfrac{1+(q-2)\tilde{x}}{e^K(z-2)}$, while infinite 1-cluster is absent when $\tilde{x} < \tilde{w}_1$, see (36). Hence the non-percolating phase exists when $\tilde{x} < \tilde{w}_1 < \dfrac{1+(q-2)\tilde{x}}{e^K(z-2)}$. This inequality needs the following condition to be fulfilled

$$\tilde{x} < \dfrac{1}{e^K(z-2)+2-q} \equiv x_{p1},$$

so the percolation transition takes place at $\tilde{x} = x_{p1}$ or at

$$H_{p1} = (z-2)J + T(z-1)\ln\dfrac{z-2+(2-q)e^{-J/T}+(q-1)e^{-2J/T}}{z-1} - T\ln\left[z-2+(2-q)e^{-J/T}\right] \quad (52)$$

In the stable thermodynamic phases $\dfrac{\partial \tilde{x}}{\partial h} = -x^{-2}\dfrac{\partial x}{\partial h} > 0$ so the non-percolating condition $\tilde{x} < x_{p1}$ means that the infinite 1-cluster is absent at $H < H_{p1}$ while it is present at $H_{p1} < H$.

Note, that when $z < q$ (52) is valid only at $T < J/\ln\dfrac{q-2}{z-2} \equiv T_1$, above this temperature the infinite 1-cluster does not exist.

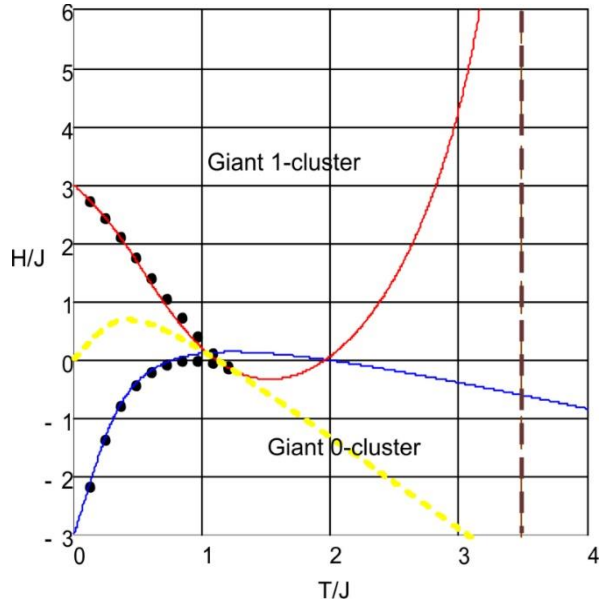

Figure 4. Phase diagram at $q < z$. $q = 6$, $z = 5$. The giant 1-cluster is absent above $T_1 = 3.476$ shown by dashed vertical line.

Here we should remind that in the region denoted as "Giant 1-cluster" the other giant $\alpha - clusters$ exist with $\alpha > 1$ due to the permutation symmetry of partition function (3).

## 5. Discussion and conclusions

We show that the calculation of specific double-Potts partition function is the useful method to study the percolation of geometric Potts clusters. With it, the percolation phase diagram and the size distribution of $\alpha - clusters$ in ferromagnetic Potts model on the Bethe lattice are found analytically. The size distribution in such correlated percolation appears to be proportional to that of the classical non-correlated bond percolation with the bond occupation probability depending on Potts model parameters ($K$, $h$) same as for the Ising clusters on this lattice [19]. Accordingly, the $\alpha - clusters$ percolation has the same classical critical indexes.

It seems that this result is not solely the property of the Bethe lattice model as the last is the good mean-field approximation for graphs and lattices with $z \gg 1$ outside the critical region. Probably, in the mean-field percolation region of a wide class of the Potts and Ising ferromagnetic models the correlations amount to the formation of independent pairs of nearest neighbor like-valued spins.

Beside providing the bond occupation probability, the influence of the model on the percolation of its geometric clusters is rather scarce. We may note that at $T = 0$ giant clusters exist strictly outside coexistence region. Meanwhile, at finite $T$ both $H_{p0}$ and $H_{p1}$ cross the line of the first order phase transition, cf. Figure 1 (b), which means that the thermodynamic transition is not a consequence of the percolation of geometric clusters.

The present approach can be useful for the numerical studies of Potts clusters' percolation on Euclidean lattices. Thus, for rough estimate of $\alpha - clusters$ size generation function the usual Monte Carlo simulations can be used to obtain $Z_{\tilde{q}}(\zeta)$ for several integer $\tilde{q}$ and to interpolate it to $\tilde{q} = 1$. To get more precise results one should extend the expression (3) for $Z_{\tilde{q}}(\zeta)$ to real $\tilde{q}$. This can be done, for example, within the transfer matrix representation of $Z_{\tilde{q}}(\zeta)$, see Ref. [22]. Note also, that present method can be easily modified for other types of Potts clusters.

## Acknowledgment

Research was financially supported by the Ministry of Science and Higher Education of the Russian Federation (State assignment in the field of scientific activity, Southern Federal University, 2020).

Appendix A

Using the relations that follow from (13)

$$V_0(0,0) = W_0(1,0)\{(q-1)e^h[1+(\tilde{q}-1)\zeta]+e^K v\},$$

$$V_0(0,1) = W_0(1,0)\{(q-1)e^h[1+(\tilde{q}-1)\zeta]+e^K \zeta u\},$$

$$V_0(1,0) = W_0(1,0)\{v+(\tilde{q}-1)\zeta u+e^h(e^K+q-2)[1+(\tilde{q}-1)\zeta]\},$$

$$V_0(1,1) = V_0(1,0), \quad W_{1,0}^{2-z} = \{v+(\tilde{q}-1)\zeta u+e^h(e^K+q-2)[1+(\tilde{q}-1)\zeta]\}^{z-1}$$

we get the following equations for $u$ and $v$

$$v = \left\{\frac{(q-1)e^h[1+(\tilde{q}-1)\zeta]+e^K v}{v+(\tilde{q}-1)\zeta u+e^h(e^K+q-2)[1+(\tilde{q}-1)\zeta]}\right\}^{z-1}, \quad u = \left\{\frac{(q-1)e^h[1+(\tilde{q}-1)\zeta]+e^K \zeta u}{v+(\tilde{q}-1)\zeta u+e^h(e^K+q-2)[1+(\tilde{q}-1)\zeta]}\right\}^{z-1} \quad (A1)$$

Then we can express potential for $\alpha = 0$ via just two variables, $u$ and $v$ as follows

$$F_{\tilde{q}}(h,\zeta,\alpha=0) = (z-1)\ln\{v+(\tilde{q}-1)\zeta u+e^h(e^K+q-2)[1+(\tilde{q}-1)\zeta]\} +$$

$$+\frac{2-z}{2}\ln\left\{\begin{array}{l}v[(q-1)e^h[1+(\tilde{q}-1)\zeta]+e^K v]+(q-1)e^h[v+(\tilde{q}-1)\zeta u+e^h(e^K+q-2)[1+(\tilde{q}-1)\zeta]]+\\+(\tilde{q}-1)\zeta[u[(q-1)e^h[1+(\tilde{q}-1)\zeta]+e^K \zeta u]+(q-1)e^h[v+(\tilde{q}-1)\zeta u+e^h(e^K+q-2)[1+(\tilde{q}-1)\zeta]]]\end{array}\right\} \quad (A2)$$

Appendix B

The equation (47) follows from the Maxwell rule [20]

$$\int_{m_P(x_1)}^{m_P(x_2)} dm\, h[x(m)] - h_{tr}[m_P(x_2) - m_P(x_1)] = 0. \quad (B1)$$

Here $x(m)$ is the inverse function to $m_P(x)$ so $h[x(m)] = h(x)$ and $h_{tr}$ is the transition field such that

$$h_{tr} = h(x_2) = h(x_1). \quad (B2)$$

Introducing the function

$$\Phi(m) = \int_{1/2}^{m} d\tilde{m}[h(\tilde{m}) - h_{tr}], \quad (B3)$$

we can represent (B1) as

$$\Phi[m_P(x_2)] = \Phi[m_P(x_1)].$$

Integrating (B3) by parts and using (B2) we have

$$\Phi[m_P(x_n)] = m_P(x_n)[h(x_n) - h_{tr}] - \frac{1}{2}\left[h\left(m=\frac{1}{2}\right) - h_{tr}\right] - \int_{h\left(m=\frac{1}{2}\right)}^{h(x_n)} d\tilde{m}\frac{\partial h(\tilde{m})}{\partial \tilde{m}} = -\frac{1}{2}\left[h\left(m=\frac{1}{2}\right) - h_{tr}\right] - \int_{h\left(m=\frac{1}{2}\right)}^{h_{tr}} d\tilde{m}\frac{\partial h(\tilde{m})}{\partial \tilde{m}}$$

Thus $\Phi[m_P(x_2)] = \Phi[m_P(x_1)] = 0$ when $h_{tr} = h\left(m = \frac{1}{2}\right)$. Hence, the first order transition takes place at the field

$$h_{tr} = h(x_{tr}) = (z-1)\ln\frac{e^K x_{tr} + q - 1}{x_{tr} + e^K + q - 2} - \ln x_{tr}$$

where $x_{tr}$ is defined through the equation $m_P(x_{tr}) = \frac{1}{2}$. We easily find that

$$x_{tr} = \sqrt{(q-1)\left[1 + (q-2)e^{-K}\right]}$$

which gives $H_{tr}(z,q,T) = Th(x_{tr})$ in (47).

REFERENCES

1. F. Y. Wu, Rev. Mod. Phys. **54** 235 (1982).

2. R. J. Baxter, *Exactly solved model in statistical mechanics*, (Academic Press, London, 1982).

3, C. Tsallis, A. C. N. de Magalhaes, Physics Reports **268** 305 (1996).

4. Hui-Jia Li, Yong Wang, Ling-Yun Wu, Junhua Zhang, Xiang-Sun Zhang, Phys. Rev. E **86**(1) 012801 (2012).

5. A. Coniglio and A. Fierro, in *Encyclopedia of Complexity and Systems Science*, Part 3, 1596 (Springer, New York, 2009); arXiv: 1609.04160.

6. C. R. Nappi, F. Peruggi and L. Russo, J. Phys. **A:** Math. Gen. **10** 205 (1977).

7. A. Coniglio, Phys. Rev. B **13** 2194 (1976).

8. K. K. Murata, J. Phys. A: Math. Gen. **12** 81 (1979).

9. A. Coniglio and W. Klein, J. Phys. A: Math. Gen. **13** 2775 (1980).

10. A. A. Saberi, J. Stat. Mech. P07030 (2009).

11. A. A. Saberi and H. Dashti-Naserabadi, Euro. Phys. Lett. **92** 67005 (2010).

12. J. D. Noh, H. K. Lee, and H. Park, Phys. Rev. E **84** 010101 (2011).

13. W. S. Jo, S. D. Yi, S. K. Baek, and B. J. Kim, Phys. Rev. E **86** 032103 (2012).

14. A. R. de la Rocha, P. M. C. de Oliveira and J. J. Arenzon, Phys. Rev. E **91** 042113 (2015).

15. H. E. Stanley, J. Phys. A: Math. Gen. **12** L211 (1979).

16. P. M. Kogut and P. L. Leath, J. Phys. C: Solid State Phys. **15** 4225 (1982);

    N. S. Branco, S. L. A. de Queiroz and R. R. dos Santos, J. Phys. C: Solid State Phys. **19** 1909 (1986).

17. C. M. Fortuin and P. W. Kasteleyn, Physica **57** 536 (1972).

18. M. J. Stephen, Phys. Rev. B **15** 5674 (1977).

19. P. N. Timonin, Physica A **527** 121402 (2019).

20. J. L. Lebowitz and O. Penrose, J. of Math. Phys. **7** 98 (1966).

21. F. Peruggi, F. di Liberto and G. Monroy, J. Phys. A: Math. Gen. **16** 811 (1983).

22. J. L. Jacobsen and J. Cardy, Nuclear Phys. B **515** 701 (1998).